\documentclass[]{spie}  %>>> use for US letter paper
\usepackage{amsmath,amsfonts,amssymb}
\usepackage{graphicx}
\usepackage[colorlinks=true, allcolors=blue]{hyperref}

\title{Gravity+ Adaptive Optics (GPAO) tests in Europe}

\author[a]{Florentin Millour}
\author[b]{Guillaume Bourdarot}
\author[c]{Jean-Baptiste Lebouquin}
\author[d]{Anthony Berdeu}
\author[a]{Mathis Houllé}
\author[a]{Philippe Berio}
\author[d]{Thibaut Paumard}
\author[e]{Denis Defrère}
\author[f]{Paulo Garcia}
\author[g]{Ferreol Soulez}
\author[h]{Sebastian Hoenig}
\author[i]{Fatmé Allouche}
\author[b]{Martin Bachbucher}
\author[a]{Christophe Bailet}
\author[d]{Cyrille Blanchard}
\author[a]{Olivier Boebion}
\author[i]{Henri Bonnet}
\author[b]{Amit Brara}
\author[a]{Marcel Carbillet}
\author[b]{Stephan Czempiel}
\author[c]{Alain Delboulbé}
\author[d]{Roderick Dembet}
\author[d]{Clémence Edouard}
\author[b]{Frank Eisenhauer}
\author[b]{Helmut Feuchtgruber}
\author[b]{Christoph Furchstsam}
\author[b]{Stefan Gillessen}
\author[b]{Armin Goldbrunner}
\author[f]{Tiago Gomes}
\author[a]{Carole Gouvret}
\author[c]{Sylvain Guieu}
\author[b]{Mike Hartl}
\author[b]{Johannes Hartwig}
\author[b]{Frank Haussmann}
\author[b]{David Huber}
\author[d]{Ihsan Ibn Taïeb}
\author[i]{Johann Kolb}
\author[a]{Stéphane Lagarde}
\author[a]{Olivier Lai}
\author[a]{James Leftley}
\author[b]{Dieter Lutz}
\author[c]{Yves Magnard}
\author[a]{Aurélie Marcotto}
\author[c]{Hugo Nowacki}
\author[i]{Sylvain Oberti}
\author[b]{Thomas Ott}
\author[b]{Christian Rau}
\author[a]{Sylvie Robbe-Dubois}
\author[a]{Jules Scigliuto}
\author[b]{Franz Soller}
\author[i]{Pavel Shchekaturov}
\author[b]{Daniel Schuppe}
\author[c]{Eric Stadler}
\author[b]{Sinem Uysal}
\author[b]{Felix Widmann}
\author[b]{Ekherhard Wieprecht}
\author[i]{Julien Woillez}
\author[b]{Şenol Yazici}
\affil[a]{Univ. C\^ote d'Azur, Observatoire de la C\^ote d'Azur, CNRS, Laboratoire Lagrange, (France)}
\affil[b]{Max Planck Institute for extraterrestrial Physics, Giessenbach-straße 1, 85748 Garching (Germany)}
\affil[c]{Univ. Grenoble Alpes, CNRS, IPAG, 38000 Grenoble (France)}
\affil[d]{LESIA, Observatoire de Paris, Univ. PSL, CNRS, Sorbonne Univ., Univ. Paris Cité, 5 place Jules Janssen, 92195 Meudon (France)}
\affil[e]{Institute of Astronomy, KU Leuven, Celestijnenlaan 200D, 3001, Leuven (Belgium)}
\affil[f]{Faculdade de Engenharia, Univ. do Porto, rua Dr. Roberto Frias, 4200-465 Porto (Portugal)}
\affil[g]{Univ. Lyon, Univ. Lyon1, Ens de Lyon, CNRS, Centre de Recherche Astrophysique de Lyon UMR5574, F-69230, Saint-Genis-Laval, (France)}
\affil[h]{Department of Physics \& Astronomy, Univ. of Southampton, Southampton, SO17 1BJ (UK)}
\affil[i]{European Southern Observatory, Karl-Schwarzschild-Straße 2, 85748 Garching (Germany)}

\authorinfo{Further author information: (Send correspondence to F. Millour)\\F. Millour: E-mail: fmillour@oca.eu, Telephone: +33 489 15 03 59}

\begin{document} 
\maketitle

\begin{abstract}
%{\bf Chat GPT abstract:}
%This paper presents the results of the GRAVITY+ Adaptive Optics (GPAO) tests conducted in Europe from 2021 to 2024. The GPAO system, designed to enhance the adaptive optics capabilities of the GRAVITY instrument, comprises a WaveFront Sensor (WFS), Corrective Optics (CO) including a Deformable Mirror (DM), and a Real-Time Computer (RTC). The testing involved assembling and integrating the GPAO subsystems at the Lagrange laboratory in Nice, simulating conditions at the Paranal Observatory. The testing phases demonstrated that the adaptive optics system meets performance expectations, leading to the approval for deployment at the Paranal Observatory. Key milestones included the integration and testing of two GPAO systems, calibration procedures, and performance evaluations under simulated turbulent conditions. The tests validated the system’s ability to achieve high Strehl ratios and robustness against various observational challenges.
%{\bf Florentin's abstract:}
We present in this proceeding the results of the test phase of the GRAVITY+ adaptive optics. This extreme AO will enable both high-dynamic range observations of faint companions (including exoplanets) thanks to a 40x40 sub-apertures wavefront control, and sensitive observations (including AGNs) thanks to the addition of a laser guide star to each UT of the VLT. This leap forward is made thanks to a mostly automated setup of the AO, including calibration of the NCPAs, that we tested in Europe on the UT+atmosphere simulator we built in Nice. We managed to reproduce in laboratory the expected performances of all the modes of the AO, including under non-optimal atmospheric or telescope alignment conditions, giving us the green light to proceed with the Assembly, Integration and Verification phase in Paranal.
\end{abstract}

\keywords{GRAVITY+, XAO, adaptive optics, AIT, test, VLTI perspectives}

\section{INTRODUCTION}
\label{sec:intro}  % \label{} allows reference to this section

We present in this paper some of the results of the GRAVITY+ Adaptive Optics (GPAO) test plan performed in Europe from 2021 to 2024, showing that the adaptive optics performs as expected. These results triggered the green light to send GPAO to the Paranal Observatory in mid-2024.

\paragraph{Short presentation of GPAO}

\begin{figure} [htbp]
   \begin{center}
   \begin{tabular}{c}
   \includegraphics[width=0.95\textwidth]{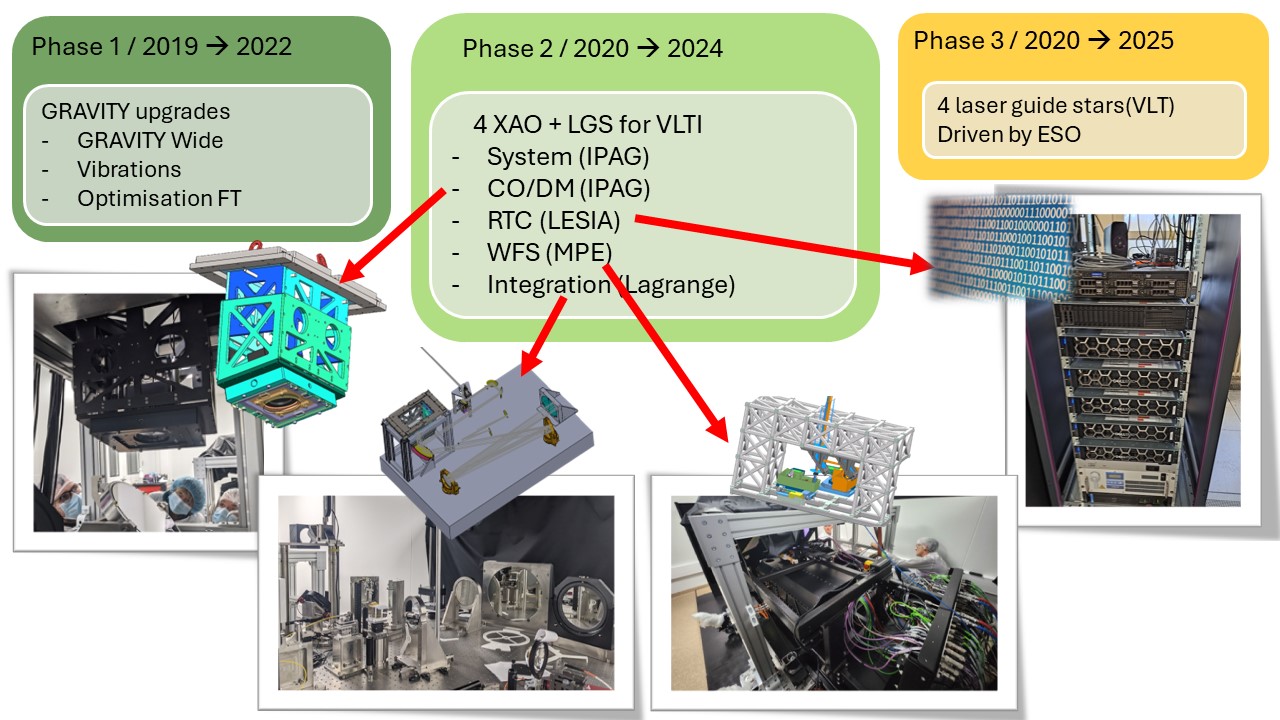}
   \end{tabular}
   \end{center}
   \caption[example] 
   { \label{fig:GRAVITY+timeline} 
Global timeline of the GRAVITY+ project. We show below the timeline CAD illustrations of the different subsystems, as well as pictures taken during the Assembly, Integration and Tests of GPAO in Nice.
}
\end{figure} 

The GRAVITY+ Adaptive Optics (GPAO) is an extreme AO that will enable both high-dynamic range observations of faint companions (including exoplanets) thanks to a 40x40 sub-apertures wavefront control, and sensitive observations (including AGNs) thanks to the addition of a laser guide star to each UT of the VLT. GPAO comprises 3 subsystems: a Wavefront Sensor (WFS)\cite{2024SPIEB}, a Corrective Optics (CO - including a Deformable Mirror DM), and a Real Time Computer (RTC). Each subsystem is assembled and tested by one consortium institute (WFS: MPE ; CO: IPAG ; and RTC: LESIA). GPAO also include the infrared 9x9 WFS from the former CIAO AO and adds to it the laser guide star, summing GPAO modes to 4: NGS-VIS (40x40), LGS-VIS (4x4 + 30x30), NGS-IR (9x9) and LGS-IR (9x9 + 30x30). 

To ensure these GPAO subsystems work together smoothly, and to measure the effective performances of the adaptive optics, we built a test bench in Nice (Lagrange laboratory) that reproduced an UT coudé focus as well as a turbulent atmosphere with typical Paranal conditions\cite{2022SPIE12183E..1XM}. GPAO is part of the whole GRAVITY+ project\cite{2022Msngr.189...17A,2019vltt.confE..30E}, which includes as well dual field interferometry\cite{2022AandA...665A..75G}, upgrades of GRAVITY\cite{2017AandA...602A..94G, 2024AandA...684A.184N} and the VLTI\cite{2022SPIE12183E..1ZB}, and laser guide stars installed on the telescopes.

\section{Timeline of the GPAO test plan}

A global timeline of the GRAVITY+ project is presented in Figure~\ref{fig:GRAVITY+timeline}. GPAO corresponds to the phase 2 of the project, phase 1 being upgrades of the instruments, and phase 3 being the introduction of a laser guide star to every UT.
The GPAO test plan timeline followed closely the Assembly and Integration ones, with activities of integration and tests conducted in parallel. This way of working was very efficient thanks to the test bench in Nice, that saw two complete GPAO systems integrated in sequence: GPAO\#1 and GPAO\#2.
The early developments were performed on the GPAO\#1 (Section~\ref{sec:GPAO1}), replaced by GPAO\#2 in a second phase (Section~\ref{sec:GPAO2}), which included the finalized version of all GPAO subsystems and was used for the qualification to Chile.

\subsection{GPAO\#1 / 2021 -- 2023}
\label{sec:GPAO1}

GPAO\#1 was used very early in the project (as early as 2021) to help progress in parallel the development of GPAO and the test plan. It was used to perform some low-level aspects of the test plan. GPAO\#1 was extremely helpful to provide already a workable system although with features slightly different from the final design. 

\begin{figure} [htbp]
   \begin{center}
   \begin{tabular}{c}
   \includegraphics[width=0.6\textwidth]{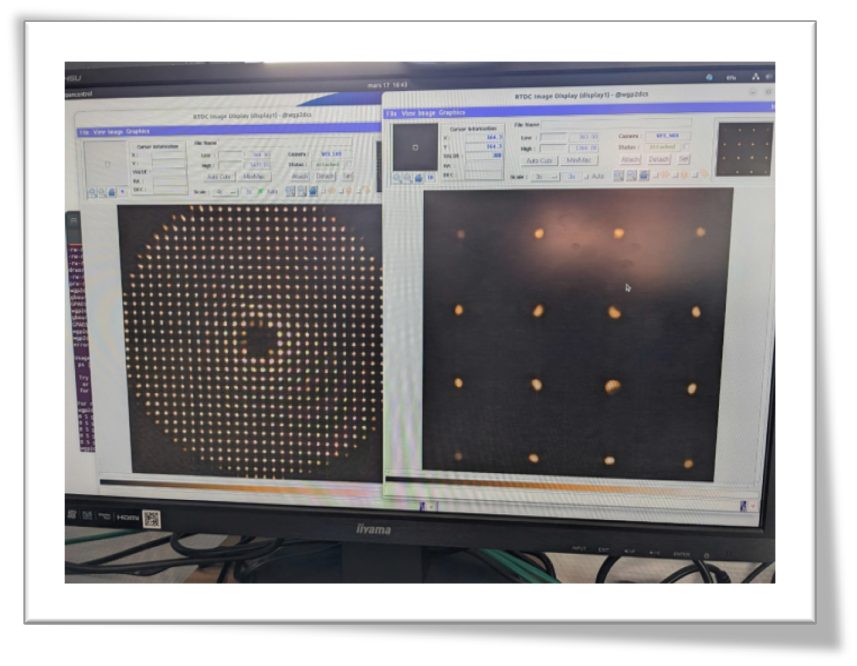}
   \includegraphics[width=0.36\textwidth]{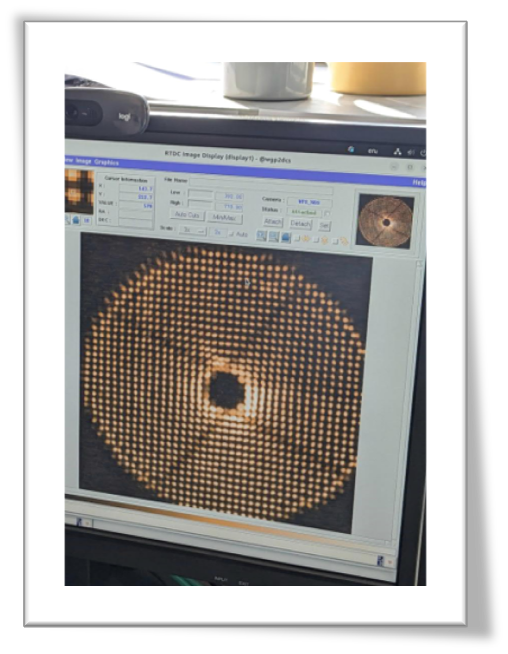}
   \end{tabular}
   \end{center}
   \caption[example] 
   { \label{fig:detector_look} 
Pictures showing two WFS setups of GPAO: On the left is the LGS mode with the laser guide star on the 30x30 SHWFS (left), while the natural guide star is acquired with a low order SHWFS (4x4, right). On the right the NGS setup is shown, with a 40x40 SHWFS on a visible natural guide star.
}
\end{figure}

GPAO\#1 WFS first included only the LGS module (30x30, see Figure~\ref{fig:detector_look}, left), to give way to a fast track integration of this most complicated part of the AO. The NGS module was added at a later step (40x40 and 4x4, see Figure~\ref{fig:detector_look}, middle and right). 
GPAO\#1 WFS was sent back to MPE for retro-fitting after GPAO\#2 WFS was installed in Nice.

The GPAO CO is composed of a gimbal and the ALPAO deformable mirror itself\cite{2014SPIE.9148E..25C}. Several tests made on GPAO\#1 have been made with a pre-series DM\#0 (BAX437) with slightly different characteristics than the final DM: limited stroke, 41x41 actuators, outer circle of actuators within the useful pupil.
The final DM\#1 (BAX651) has been added to GPAO\#1 at a later stage to test the full AO, together with an upgraded CO (to accommodate the increased size of the DM).

The RTC is composed of a cluster of 4 machines (the soft [S]RTC) for telemetry recording and matrices update, and a fast and optimized machine for matrix multiplication (the hard [H]RTC). Additional computers are installed in the rack : the workstation computer, and a logger used to measure the loop latency.
It is housed in an electronic cabinet. We first used a standard computer cabinet to host the RTC computers.

GPAO\#1 had several limitations which we list here:
\begin{itemize}
    \item No dichroic plate at the WFS entrance (replaced by a 50/50 beamsplitter),
    \item Pico-motors (Newport) were initially used for the Pupil Steering Mirror, in replacement of Piezo-actuators (due to a hardware shortage), which were implemented on GPAO\#2,
    \item Gain and Electronic instabilities on the OCAM2K cameras, which were fixed at a later stage,
    \item Various time-constrained activities that prevented from fulfilling the full set of tests on GPAO\#1, especially for the CIAO modes.
\end{itemize}

GPAO\#1, as well as GPAO\#3 and 4 have been sent as separated subsystems from each institute in charge (MPE, IPAG, MPE) to ESO for a shipment to Paranal at the end of June 2024.

\subsection{GPAO\#2 / 2023 -- 2024}
\label{sec:GPAO2}

GPAO\#2 was the final version of GPAO, including the final versions of the WFS (WFS\#2), DM (both DM\#1 - BAX651 - and DM\#2 - BAX652 - have been tested on GPAO\#2). The RTC was upgraded with its final informatics cabinet during the integration of GPAO\#2. All the different GPAO modes (40x40, 30x30 + 4x4, 9x9 and 30x30 + 9x9) were tested until June 2024,
and sent to ESO directly from Nice.
Most of the tests presented in this paper were conducted on GPAO\#2.

\begin{figure} [htbp]
   \begin{center}
   \begin{tabular}{c}
   \includegraphics[width=0.26\textwidth]{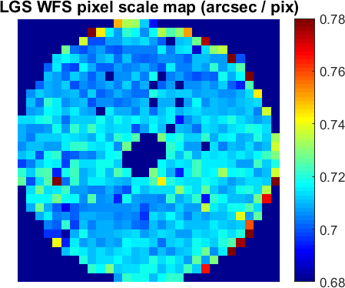}
   \includegraphics[width=0.26\textwidth]{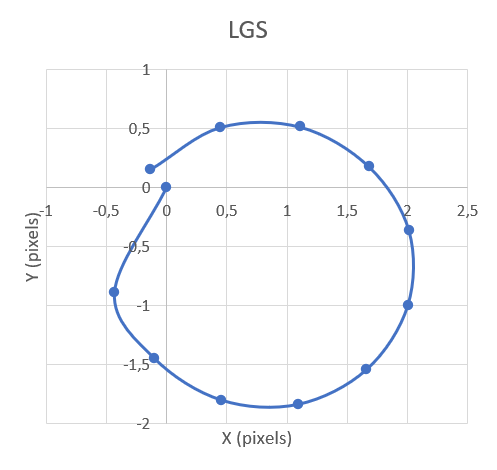}
   \includegraphics[width=0.5\textwidth]{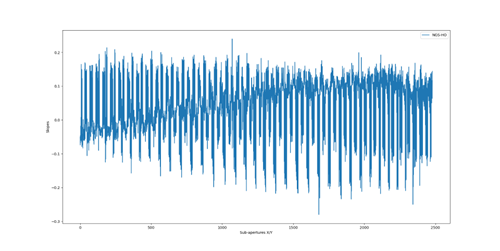}
   \end{tabular}
   \end{center}
   \caption[example] 
   { \label{fig:GPAO_characteristics} 
Some of the tested aspects of GPAO that are used as calibrations in the system operation. {\bf Left:} Pixel scale for each sub-aperture in the LGS WFS. {\bf Middle:} Derotator wobble in WFS pixels, showing a very good optical alignment. {\bf Right:} Reference slopes vector measured using the GUI panels of GPAO.
}
\end{figure} 

\section{Basic tests}

To have the GPAO system up and running, we developed several templates (template = daytime script executed on the telescope console, compatible with the VLTI software, see Table~\ref{tab:templates}) to calibrate many of the low level (camera checks, motors calibrations, lookup tables, pixel scales) and high level elements of GPAO (matrices of GPAO, reference slopes, NCPA, performances). This allowed us to use these templates for the tests, that were easy to repeat along the system development (that was done in parallel to the test plan). We show in the next sections some of the key calibrations that are used in GPAO.

\subsection{Cameras check}

The high framerate and low noise OCAM2K cameras of GPAO are delicate and fragile, therefore their detector health must be monitored regularly. We developed a complete health check monitoring template \texttt{gen\_tec\_ocam\_check} to survey the EMCCD gain, detector noise, octants health in a convenient pdf report that is generated automatically.

\subsection{pixel scale}

The pixel scale measurement (see Figure~\ref{fig:GPAO_characteristics}) is included in the GPAO template \texttt{gen\_tec\_pixel\_scale}. It simply scans the XY table positions in both X- and Y-axes and measures the slopes (Positions of spots barycentre) through the WFS. We find a pixel scale of about 0.73"/pixel, which is close but not identical to the design pixel scale of 0.8’’/px. This has an impact on the later steps (especially matrix generation) and must be taken into account.

\begin{table} [htbp]
    \centering
    \begin{tabular}{r|c|l}
        GPAO template name  & Description                     & Output \\
        \hline
        \texttt{gen\_tec\_ocam\_check}  & Verify OCAM2K characteristics     & OCAM2K health check report\\
        \texttt{gen\_tec\_pixel\_scale} & Measure WFS pixel scales        & Pixel scale\\
        \texttt{gen\_tec\_psm\_matrix}  & Matrix PSM-WFS                  & Calibration matrix\\
        \texttt{gen\_tec\_xyt\_matrix}  & Matrix XYT-WFS                  & Calibration matrix\\
        \texttt{gen\_tec\_qsm\_matrix}  & Matrix QSM-WFS                  & Calibration matrix\\
        \texttt{gen\_tec\_adc\_lut}     & Lookup table for ADC            & Lookup table\\
        \texttt{gen\_tec\_rot\_lut}     & Lookup table for derotator      & Lookup table\\
        \texttt{gen\_tec\_ao\_matrices} & Produces all useful AO matrices & Set of SPARTA matrices\\
        \texttt{gen\_tec\_ref\_slopes}  & Measures the reference slopes   & Reference slopes\\
        \texttt{gen\_tec\_ref\_ncpa}    & Measure and apply NCPAs         & Modal vector of NCPA\\
        \hline
    \end{tabular}
    \caption{Set of templates (daytime telescope scripts) used to calibrate GPAO and set it ready to close the loops and get a high strehl ratio.}
    \label{tab:templates}
\end{table}

\subsection{Motors interaction matrices}

Several motors matrices have to be calibrated within GPAO: QSM-WFS (CO tip/tilt vs WFS), XYT-WFS (XY vs WFS, the XY table allows field selection in the AO), and DM-WFS (DM tip/tilt vs WFS).

These matrices are measured with dedicated templates: \texttt{gen\_tec\_qsm\_matrix}, \texttt{gen\_tec\_xyt\_matrix}, and \texttt{gen\_tec\_dm\_matrix}. Feeding these matrices to the GPAO OS allows us to input directly sky coordinates as offsets in the system.

\begin{figure}[htbp]
   \begin{center}
   \begin{tabular}{c}
   \includegraphics[width=0.28\textwidth]{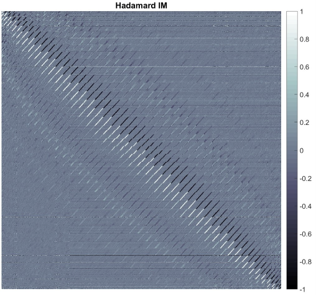}
   \includegraphics[width=0.28\textwidth]{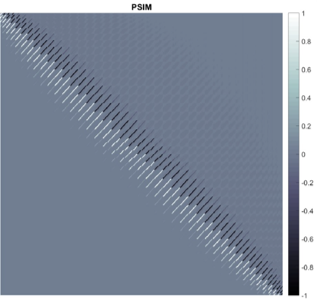}
   \includegraphics[width=0.42\textwidth]{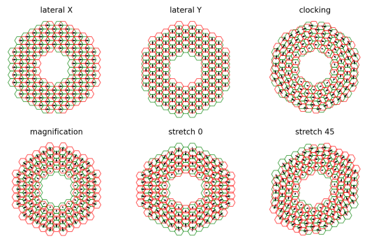}
   \end{tabular}
   \end{center}
   \caption[example] 
   { \label{fig:PSIM_vs_real_life} 
{\bf Left:} A real-life interaction matrix (IM) measured on the GPAO system using Hadamard method. {\bf Middle:} Pseudo-Synthetic Interaction Matrix (PSIM) computer-generated using set parameters according to the measured characteristics of the GPAO system. Note the absence of noise. {\bf Right:} The mis-registrations that are calculated with the new algorithm of GPAO.
}
\end{figure} 

\subsection{Motors lookup tables}

In addition, we need to produce lookup tables (see Figure~\ref{fig:GPAO_characteristics}) for the derotator and the ADC (to compensate pupil and image tilt when operating them). As for motors matrices, we developed dedicated calibration templates \texttt{gen\_tec\_xyt\_rot\_lut} and \texttt{gen\_tec\_adc\_wobble} to feed the adaptive optics OS.
All the implemented calibration templates are included in a global GPAO health check template.

\subsection{Matrices generation}

We implemented a complete matrices generation set using one template \texttt{gen\_tec\_ao\_matrices}, in order to populate properly the SPARTA database that is used to control GPAO. This template uses all the previously-calibrated knowledge of the system to generate synthetic maps for the system, notably interaction matrices (IM) that are used in the AO main control loop. 
We tested the generated matrices are indeed nearly identical to measured ones directly on the system using the graphical interface (see Figure~\ref{fig:PSIM_vs_real_life}), and allows us to close the GPAO loop in a blink after setup while correcting a high number of modes (up to 800, the baseline being 500).

\subsection{Reference slopes}

The reference slopes (before NCPA correction) of both WFS have been defined as the "central" positions of each sub-aperture, i.e. position of spots with a flat DM. It is a purely geometrical definition. 
The reference slopes are measured using the \texttt{gen\_tec\_ref\_slopes} template. An example reference slopes vector is shown in Figure~\ref{fig:GPAO_characteristics}.

\begin{figure} [htbp]
   \begin{center}
   \begin{tabular}{c}
   \includegraphics[width=0.48\textwidth]{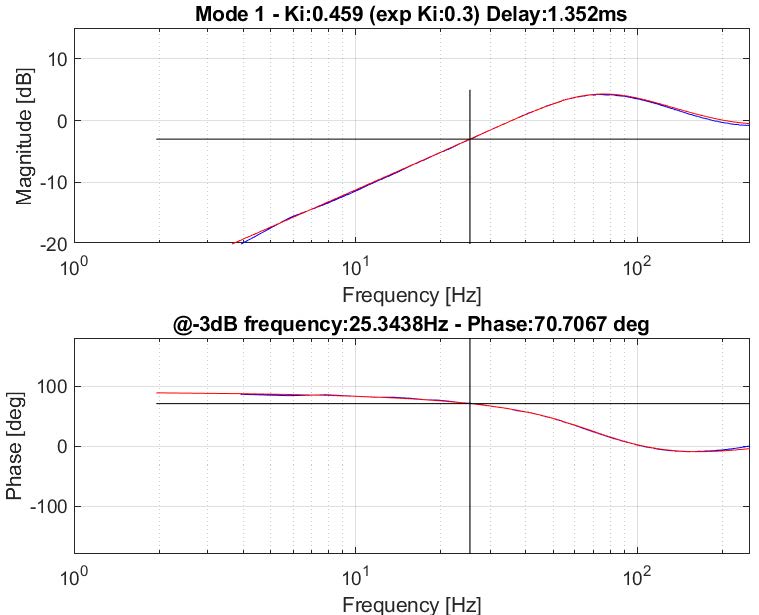}
   \includegraphics[width=0.48\textwidth]{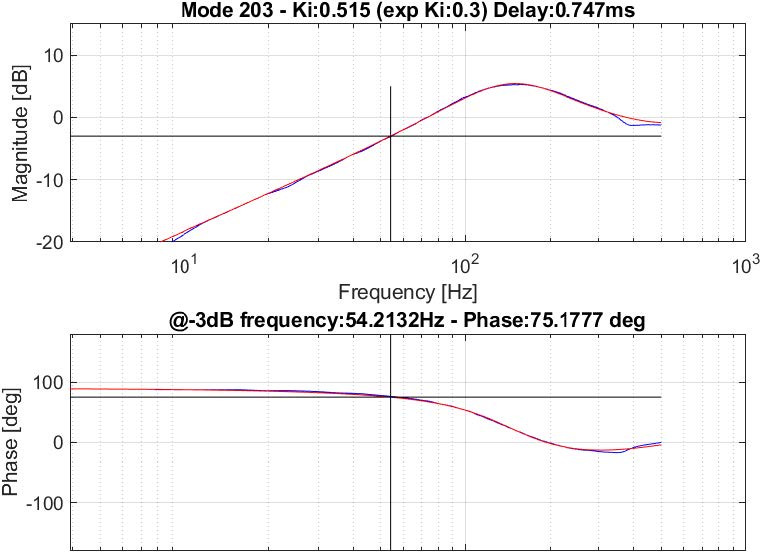}
   \end{tabular}
   \end{center}
   \caption[example] 
   { \label{fig:rejectionFunc} 
A set of measured rejection functions for GPAO and selected modes: Tip (Mode 1) and a higher order mode (Mode 203). The blue line are measurements while the red line is a rejection function model fitted to the data. This allowed us to compute effective gain (Ki) and pure delay of the loop.
}
\end{figure} 

\section{Checking GPAO}

\subsection{Functional system test of GPAO}

Now that one can calibrate the full GPAO system, we could focus on the tests, robustness and operations as well as the performances of the AO. 
To do so, we measured rejection functions on the Nice test bench for a set of corrected modes (see a few examples in Figure~\ref{fig:rejectionFunc}), and we also tested the many secondary loops that allows GPAO to keep the loop closed with changing conditions and telescope behavior (flexures, rotations, moving aberrations). To do so, we implemented and tested a new fast mis-registration algorithm \cite{2024arXiv240707423B} that tracks the WFS/DM mis-registrations in real time (Figure~\ref{fig:PSIM_vs_real_life}).

These tests enabled us to validate the baseline behavior of GPAO, with a NGS loop working nominally at 1\,kHz and controlling up to 800 modes, an LGS+LO loop working at 1\,kHz+500\,Hz, separating low order modes (tip/tilt + focus) on the natural guide star and high order modes on the LGS, a system delay below 1\,ms, and the possibility to get rid of the derotator with command matrix updates on the fly.

We tested as well the robustness of the GPAO loop to off-centered pupil (that could arise if there is vignetting), defective DM actuators, target wandering, and many other hiccups that can happen during observations and affect strehl ratio if not handled properly.

In addition, the use of the rotating phase plate\cite{2022SPIE12183E..1XM} in the Nice test bench allowed us to test the robustness of the loop against turbulence (the phase plate produced a 0.7" Seeing), and to characterize the performances of the system in a realistic environment for operations. Additional rotating phase plates that were placed in the light path of the bench allowed us to stress-test the system up to an estimated 1.4" seeing (including strong scintillation).

\subsection{Optimizing the GPAO performances}

\begin{figure} [htbp]
   \begin{center}
   \begin{tabular}{c}
   \includegraphics[width=0.6\textwidth]{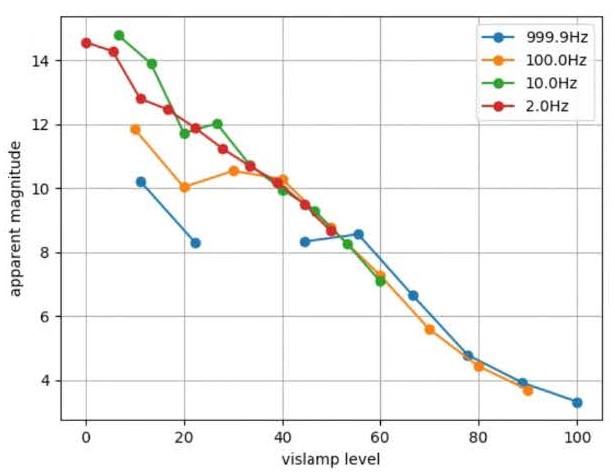}
   \includegraphics[width=0.35\textwidth]{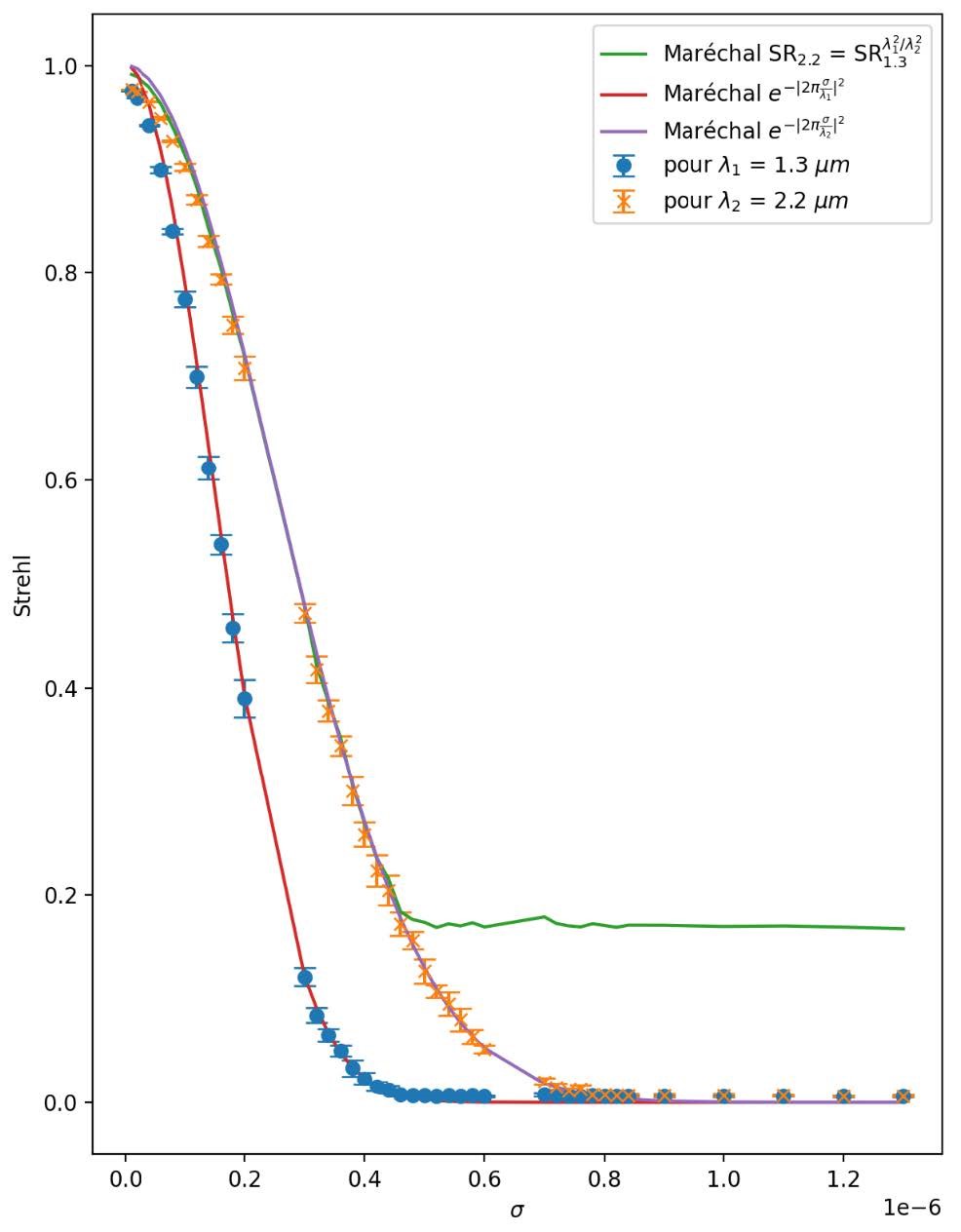}
   \end{tabular}
   \end{center}
   \caption[example] 
   { \label{fig:mag_monitor}
{\bf Left:} Bench source magnitude calibration for the 40x40 WFS. {\bf Right:} Illustration of the Marechal approximation bias at low strehl ratio, showing the issue we have to extrapolate strehl ratio from 1.3 microns (measurement wavelength, blue) to 2.2 microns (GRAVITY wavelength, green).
}
\end{figure}

We used the known characteristics of the NGS GPAO WFS to compute apparent fluxes from the bench source, in order to calibrate it in apparent magnitude for the NGS-VIS mode (Figure~\ref{fig:mag_monitor}). The low order 4x4 NGS WFS used in the LGS mode has 100x less apertures as the 40x40 NGS WFS, therefore, we simply shift this curve by 5 magnitudes to get the apparent flux in LGS mode.
The LGS source is set to a constant brightness as we expect the GPAO laser to have a roughly constant apparent magnitude.

Optimizing the GPAO loop was done for all the 4 available modes, i.e.  NGS-VIS, LGS-VIS, NGS-IR and LGS-IR. With the phase plate rotating in the bench, we varied the following parameters, for a set of input magnitudes (from 2 to 15 for NGS, and from 10 to 18 for LGS):

\begin{tabular}{ll}
    $\bullet$ Loop gain &
    $\bullet$ Loop leaks\\
    $\bullet$ Number of modes corrected&
    $\bullet$ Frame rate,\\
    $\bullet$ Pixel flux threshold&
    $\bullet$ Subaperture flux threshold\\
    $\bullet$ Use of weighting maps (or not)&
    $\bullet$ Weighting map half width half max
\end{tabular}

This optimization provided us with typical values for the GPAO loop depending on the input magnitude.
   
\section{More advanced tests}

\subsection{NCPA}

Once we set the GPAO ideal loop parameters, we started to use the bench infrared camera to measure turbulent PSFs and compute strehl ratios. A very simple way of computing the strehl S from one image can be done by measuring the brightest pixel flux in the strehl camera images:

\begin{equation}
S = S_0 \times \max_{\rm pix}({\rm PSF}_{\rm measured})
\end{equation}

One can see that this strehl measurement is multiplied by an unknown initial strehl $S_0$, but it is not an issue for the NCPA computation as we only work in relative strehl/flux: comparing the strehl S from one image to another can be simply done by measuring variation of such a flux, and a maximum flux is reached when the strehl ratio is maximum. This method is fast (not computing intensive) and can be adapted to flux-measurement devices (such as after a spatial filtering like a fiber or a pinhole), therefore it was preferred over more complete methods based on actual strehl measurements (see next section).

\begin{figure} [htbp]
   \begin{center}
   \begin{tabular}{c}
   \includegraphics[width=0.55\textwidth]{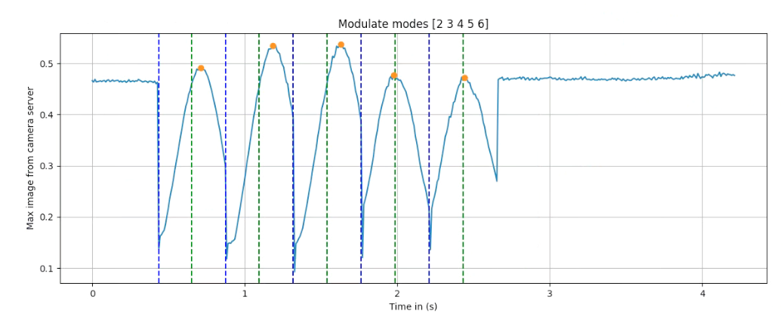}
   \includegraphics[width=0.22\textwidth]{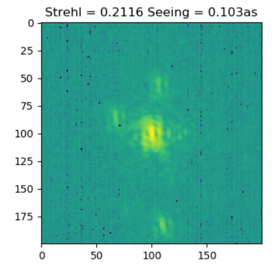}
   \includegraphics[width=0.22\textwidth]{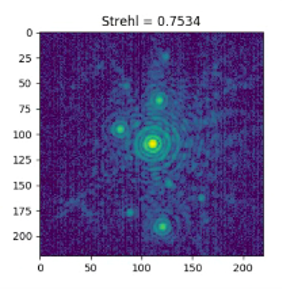}
   \end{tabular}
   \end{center}
   \caption[example] 
   { \label{fig:NCPA} 
NCPA measurements principle on the Nice bench. {\bf Left:} Flux measurement on the 1.3 microns bench strehl camera. The bell-shaped curves correspond each to one modulated mode (focus, curvature, trefoil, etc.). {\bf Middle:} Test bench PSF measured on the Strehl camera. One can see copious amount of astigmatism and spherical aberration (related to the Strehl camera position in the bench -- after two transmission plates), that translate into the low measured SR. Additional sources are parasitic reflections in the optical train in front of the SR camera. {\bf Right:} PSF measured on bench after applying NCPAs.
}
\end{figure}

The method to calibrate the NCPAs on the Nice test bench is therefore the following: we modulate one by one system modes (Zernikes or KL modes) and find the maximum flux measured with the strehl camera. Once found the maximum, we apply the modal offset and continue to the next mode. This very simple algorithm converges in about 10\,mn and produces 1.31$\mu$m strehl ratios in excess of 75\,\%, as illustrated in Figure~\ref{fig:NCPA} 

\subsection{Performances measurements}

Another aspect we tested on GPAO is its overall performances (i.e. strehl ratio as a function of apparent magnitude). We needed an effective and sufficiently accurate way of measuring the strehl S from the strehl camera images. We did it following this equation:

\begin{equation}
S = \frac{\sum_{\rm pix}{\rm PSF}_{\rm perfect, norm}}{\sum_{\rm pix}{\rm PSF}_{\rm meas., norm}} \times \frac{1}{1 - {\rm Gh}}
\end{equation}

with ${\rm PSF}_{\rm meas., norm}$ the input PSF image from the camera, cleaned up from cosmetics (in our case, a bias subtraction and line defects subtraction -- computing the median of the image line by line -- was enough), ${\rm PSF}_{\rm perfect, norm}$ a computer-generated PSF using pupil shape inputs (without the spider, but including the central obstruction), and ${\rm Gh}$ the flux fraction of ghosts in the image (that can be seen in Figure~\ref{fig:NCPA} and that account in our case to 16\% of the total flux). This strehl is measured in our case at the wavelength of the IR source, i.e. $\lambda_0 = 1.31\,\mu$m. We used the Marechal approximation to convert this strehl value to the GRAVITY wavelength where the specifications have been set up ($\lambda_1 = 2.2\,\mu$m):

\begin{equation}
S = e^{-(2\pi\sigma_\delta/\lambda)^2} \ \Rightarrow \ S(\lambda_1) = S(\lambda_0) ^ {(\lambda_0/\lambda_1)^2}
\end{equation}

We note here that 1\% of strehl at 1.3 micron (the floor precision of our strehl estimate method) translates into $\approx20$\% of strehl at 2.2 microns. The Marechal approximation does not function at these levels of strehl, as illustrated in Figure~\ref{fig:mag_monitor}, right panel, therefore all values close to 20\%
strehl in K band are subject to doubt and could be lower in practice.

The GPAO top level requirements are as follows: for median seeing conditions (0.83’’ seeing and 9.5 m/s wind), we expect a Strehl ratio of 75\% at magnitude 9 in NGS, and 50\% at
magnitude 17 in LGS. An additional limit was added at a later stage in the project: 50\% SR in K at magR = 12.5.

\begin{figure} [htbp]
   \begin{center}
   \begin{tabular}{c}
   \includegraphics[width=0.48\textwidth]{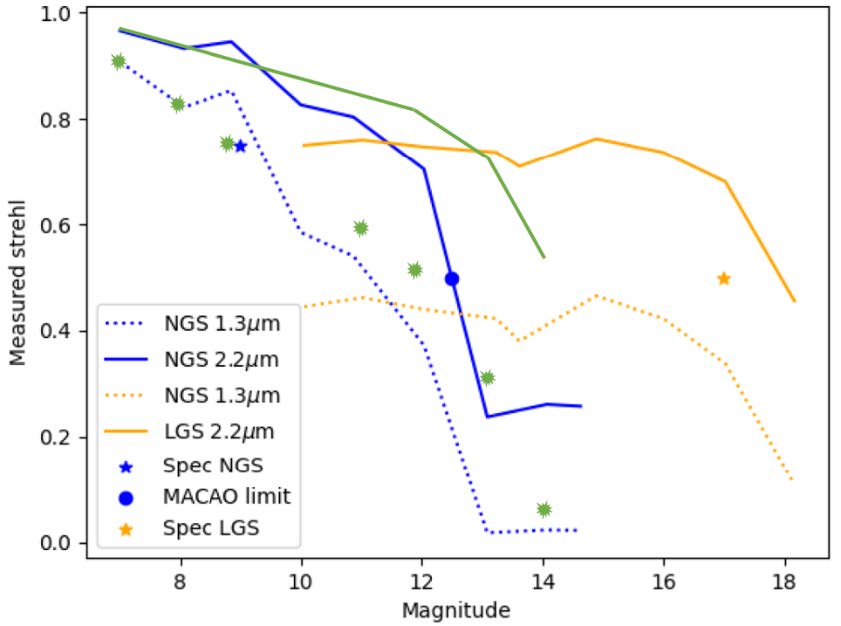}
   \includegraphics[width=0.48\textwidth, height=0.36\textwidth]{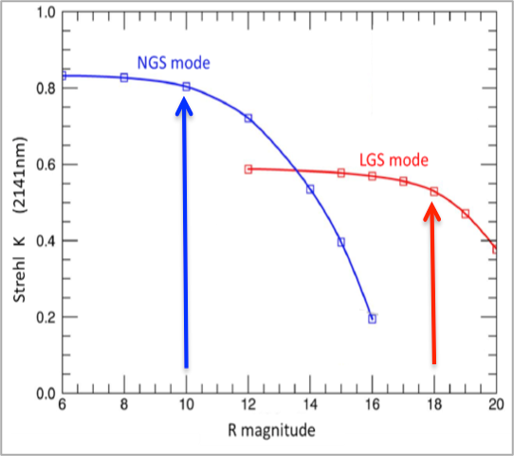}
   \end{tabular}
   \end{center}
   \caption[example] 
   { \label{fig:SR_vs_mag} 
{\bf Left:} Strehl ratios measured on the Nice bench with the rotating phase plate active. The green and blue curves are NGS measurements (green: after baffling and ESO optimization), and the yellow curve is for LGS. {\bf Right:} Expected strehl ratios on sky of GPAO. These curves are derived from the ERIS design documentation. Top level requirements are: SR=0.75 at magR=9 for NGS, and SR=0.5 at magR=17 for LGS. 
}
\end{figure}

To measure the GPAO performances, we used the flux-calibrated bench lamp at several magnitudes and measured the strehl using the above method (strehl measurement + Marechal conversion from 1.3 to 2.2\,$\mu$m), optimized GPAO loop parameters, as well as the rotating phase plate to simulate turbulence.
The resulting curves of strehl as a function of apparent magnitudes is shown in the left panel of Figure~\ref{fig:SR_vs_mag}. 
The expected values are shown in the same figure, right panel. 
The blue curve shows the GPAO performances for the NGS-VIS mode, that were just meeting the expectations with no margin. We found out that an IR LED from a motor encoder was shining photons at the OCAM2K\cite{2010SPIE.7736E..0ZF} from the NGS WFS. A dedicated baffling was designed and installed in the NGS module, which blocks this parastic light. Once the baffle in place, the measurement were repeated (green curve), showing excellent performances and meeting the expectations with some margins. The LGS performances are excellent (yellow curve), though not all effects we will see on sky could be reproduced on the bench (especially the cone effect), and a proper characterisation on sky will need to be done during the AIV and commissioning of GPAO.

\section{Conclusion and perspectives}

The Assembly, Integration and Tests of GPAO in Nice faced several major challenges, that the whole team tackled very efficiently. Among difficulties were travel restrictions, especially at the beginning of the test plan, hardware shortages, that prevented some devices to be delivered on time, and other issues with hardware. Nevertheless, the GPAO AIT team worked with a very good spirit. This allowed us to manage the challenge and deliver a system that was at Acceptance in Europe level in June 2024.

%\begin{figure} [htbp]
%   \begin{center}
%   \begin{tabular}{c}
%   \includegraphics[width=0.32\textwidth]{groupFrenchies.jpg}
%   \includegraphics[width=0.32\textwidth]{groupNCPA.jpg}
%   \includegraphics[width=0.32\textwidth]{group_MPE_Porto.jpg}\\
%   \includegraphics[width=0.32\textwidth]{group_GPAO_byebye.jpg}
%   \includegraphics[width=0.32\textwidth]{groupWithPI.jpg}
%   \includegraphics[width=0.32\textwidth]{GPAO_byebye.jpg}
%   \end{tabular}
%   \end{center}
%   \caption[example] 
%   { \label{fig:group_pictures} 
%Group pictures during the test period of GPAO, showing the nice working mood of the GPAO team. Last picture is the departure of GPAO\#2 from Nice to ESO.
%}
%\end{figure} 

Today, GPAO is flying on its way to Paranal, the Assembly, Integration and Verification phase being started from the beginning of July 2024. The future looks bright with expected transforming  changes in the VLTI performances (sensitivity\cite{2024Natur.627..281A, 2024AandA...686A.204L} and dynamic range\cite{2024AandA...686A.258P}) thanks in part to the GPAO up-to-date performances. This momentum on VLTI with new AOs\cite{2020SPIE11448E..71P}, and improved facility performances, calls for a step forward in the future, opening again perspectives on an array extension. Ideas start flourishing\cite{2024arXiv240607030L, 2024SPIEB2, 2024SF2AB} with bright goals of additional telescopes and longer, up to kilometer(s) baselines.

\begin{figure} [htbp]
   \begin{center}
   \begin{tabular}{c}
   \includegraphics[width=0.6\textwidth]{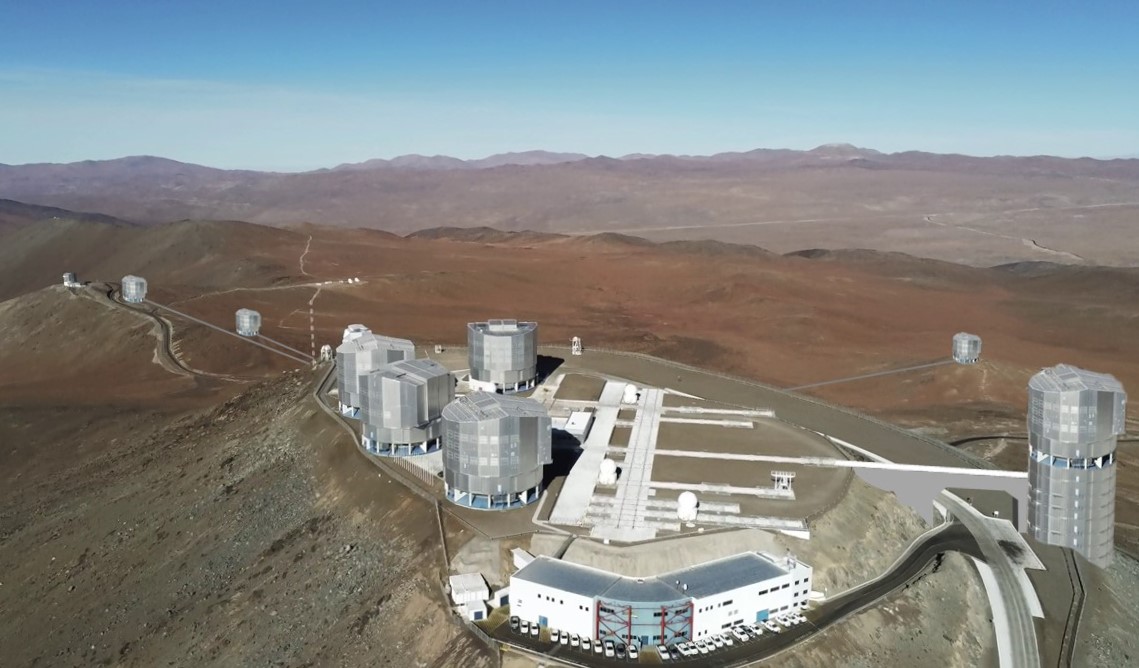}
   \includegraphics[width=0.39\textwidth]{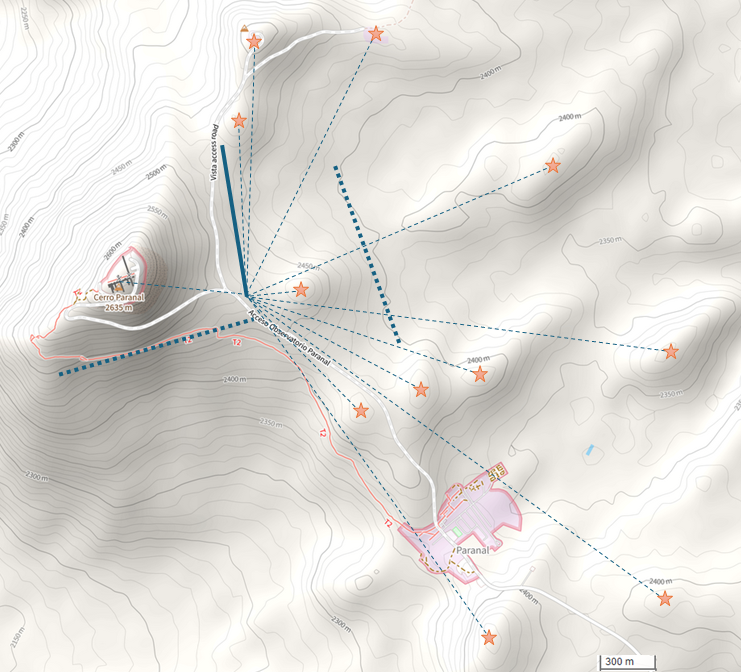}
   \end{tabular}
   \end{center}
   \caption[example] 
   { \label{fig:VLTI_extension} 
Preparing for the future. {\bf Left:} A montage showing how the VLTI may look like in 20 years from now: new UTs,  baselines longer than 1\,km and sensitivity, as well as imaging capabilities (relatively short baselines, optimised (u,v) coverage). {\bf Right:} VLT map showing a possible expansion to kilometric baselines. Orange stars mark possible new telescopes positions on local summits (to avoid shadowing). The solid line shows a possible place for a 1\,km-long delay line building (following topographic level lines, alternate places in dashed lines). The thin dashed lines show direct view from the possible telescopes sites to the DL building (including the 4 to 6 already-cophased beams of the current VLTI).
}
\end{figure}

The Paranal site is actually suited for such an horizon, as it has already the infrastructures, but also several small summits where additional telescopes could be placed, and several possible places where large delay lines buildings could be installed (Figure~\ref{fig:VLTI_extension}).

\acknowledgments % equivalent to \section*{ACKNOWLEDGMENTS}

GRAVITY+ is  a consortium composed of German (MPE, MPIA, University of Cologne), French (CNRS-INSU: LESIA, Paris ; IPAG, Grenoble ; Lagrange, Nice ; CRAL, Lyon), British (University of Southampton), Belgian (KU Leuven), and Portugese (CAUP) institutes. The instrument was built in close collaboration with ESO.
We thank Observatoire de la Côte d'Azur and Université Côte d'Azur that provided resources to manufacture and operate the GPAO testbench. This work was supported by \emph{Agence Nationale de la Recherche} (ANR) through the EXOVLTI (ANR-21-CE31-0017),  MASSIF (ANR-21-CE31-0018) and AGN MELBA (ANR-21-CE31-0011) projects. French partners thank the specific action ASHRA and the national programs PNP, PNPS and PNCG. The project was as well funded by Observatoire de Paris and Université Grenoble Alpes.
DD has received funding from the European Research Council (ERC) under the European Union's Horizon 2020 research and innovation program (grant agreement CoG - 866070). This project has received funding from the European Union's Horizon 2020 research and innovation programme under grant agreement No 101004719.

% References
\bibliography{report} % bibliography data in report.bib
\bibliographystyle{spiebib} % makes bibtex use spiebib.bst

\end{document}